\documentclass{webofc}
\usepackage[varg]{txfonts}   

\renewcommand\a{\alpha}
\renewcommand\b{\beta}
\renewcommand\d{\delta}

\renewcommand\t{\tau}

\renewcommand\c{\chi}
\renewcommand\j{\psi}
\renewcommand\o{\omega}

\newcommand\g{\gamma}

\newcommand\m{\mu}

\newcommand\p{\pi}
\newcommand\h{\theta}
\newcommand\s{\sigma}
\newcommand\f{\phi}
\newcommand\w{\eta}

\renewcommand\L{\Lambda}

\renewcommand\O{\Omega}

\newcommand\D{\Delta}

\newcommand\J{\Psi}

\newcommand{\fig}[1]{Fig.~\ref{#1}}
\newcommand{\eq}[1]{Eq.~(\ref{#1})}

\newcommand{\lan}{\langle}
\newcommand{\ran}{\rangle}

\newcommand\pt{\partial}

\newcommand{\Tr}{{\rm Tr}}

\newcommand{\br}{{\vec r}}

\newcommand{\bB}{{\vec B}}
\newcommand{\bE}{{\vec E}}
\newcommand{\bJ}{{\vec J}}

\newcommand{\jb}{{\bar \j}}

\newcommand{\rp}{{\rm RP}}

\renewcommand{\part}{{\rm part}}

\begin{document}
\title{Phenomenology of anomalous chiral transports in heavy-ion collisions}
%
%

\author{\firstname{Xu-Guang} \lastname{Huang}\inst{1,2}
}

\institute{Physics Department and Center for Particle Physics and Field Theory, Fudan University, Shanghai 200433, China
\and
           Key Laboratory of Nuclear Physics and Ion-beam Application (MOE), Fudan University, Shanghai 200433, China
          }

\abstract{%

High-energy Heavy-ion collisions can generate extremely hot quark-gluon matter and also extremely strong magnetic fields and fluid vorticity. Once coupled to chiral anomaly, the magnetic fields and fluid vorticity can induce a variety of novel transport phenomena, including the chiral magnetic effect, chiral vortical effect, etc. Some of them require the environmental violation of parity and thus provide a means to test the possible parity violation in hot strongly interacting matter. We will discuss the underlying mechanism and implications of these anomalous chiral transports in heavy-ion collisions.

}
\maketitle
\section{Introduction}
\label{intro}
High-energy heavy-ion collisions can generate deconfined quark-gluon plasma (QGP). In addition, strong magnetic fields can also be created in, e.g., non-central heavy-ion collisions~\cite{Skokov:2009qp,Deng:2012pc}. This is because that the two colliding nuclei generate two electric currents in opposite directions and thus produce a magnetic field perpendicular to the reaction plane which is the plane defined by the impact parameter and the beam direction. Strong electric fields can also be generated owing to event-by-event fluctuations or in asymmetric collisions like Cu + Au collision~\cite{Hirono:2012rt,Deng:2014uja}. Similarly, in a non-central heavy-ion collisions, a finite angular momentum can be retained in the collision region which can induce a finite vorticity. Recent numerical simulations and experimental measurement of the $\L$ hyperon spin polarization showed that such generated vorticity can be very large~\cite{Deng:2016gyh,Jiang:2016woz,STAR:2017ckg}. Thus the heavy-ion collisions provide a unique terrestrial environment to study QCD matter in strong electromagnetic field and under fast (local) rotation. We will in the next section discuss the generation of the magnetic field and flow vortcity in heavy-ion collisions.

Once coupled to the chiral anomaly, the electromagnetic field and flow vorticity can induce a number of novel transport phenomena --- the anomalous chiral transports (ACTs). One famous example is the chiral magnetic effect (CME) which is the generation of an electric current along the direction of the external magnetic field in a matter in which the right-handed fermions and left-handed fermions have unequal chemical potentials~\cite{Kharzeev:2007jp,Fukushima:2008xe}. Other ACTs include the chiral separation effect (CSE)~\cite{Son:2004tq}, chiral vortical effects (CVEs)~\cite{Erdmenger:2008rm,Banerjee:2008th,Son:2009tf}, chiral electric separation effects (CESE)~\cite{Huang:2013iia,Jiang:2014ura}, etc. New collective modes can arise through the interplay among these ACTs, e.g., the chiral magnetic wave (CMW)~\cite{Kharzeev:2010gd} and chial vortical wave (CVW)~\cite{Jiang:2015cva}. Recently, the experiments performed at RHIC and LHC generated data with features consistent with the expectation of the CME and CVE as well as the CMW. However, there exist elliptic-flow induced background effects which strongly mask the signals and make the interpretation of the experimental results ambiguous. We will give a pedagogical introduction to the underlying mechanism of the ACTs and then discuss their experimental searches.

\section{Magnetic fields and vorticity in heavy-ion collisions}
\label{sec-mag-vor}
\subsection {Magnetic fields}\label{sec:mag}
\begin{figure}
\begin{center}
\includegraphics[width=6.0cm]{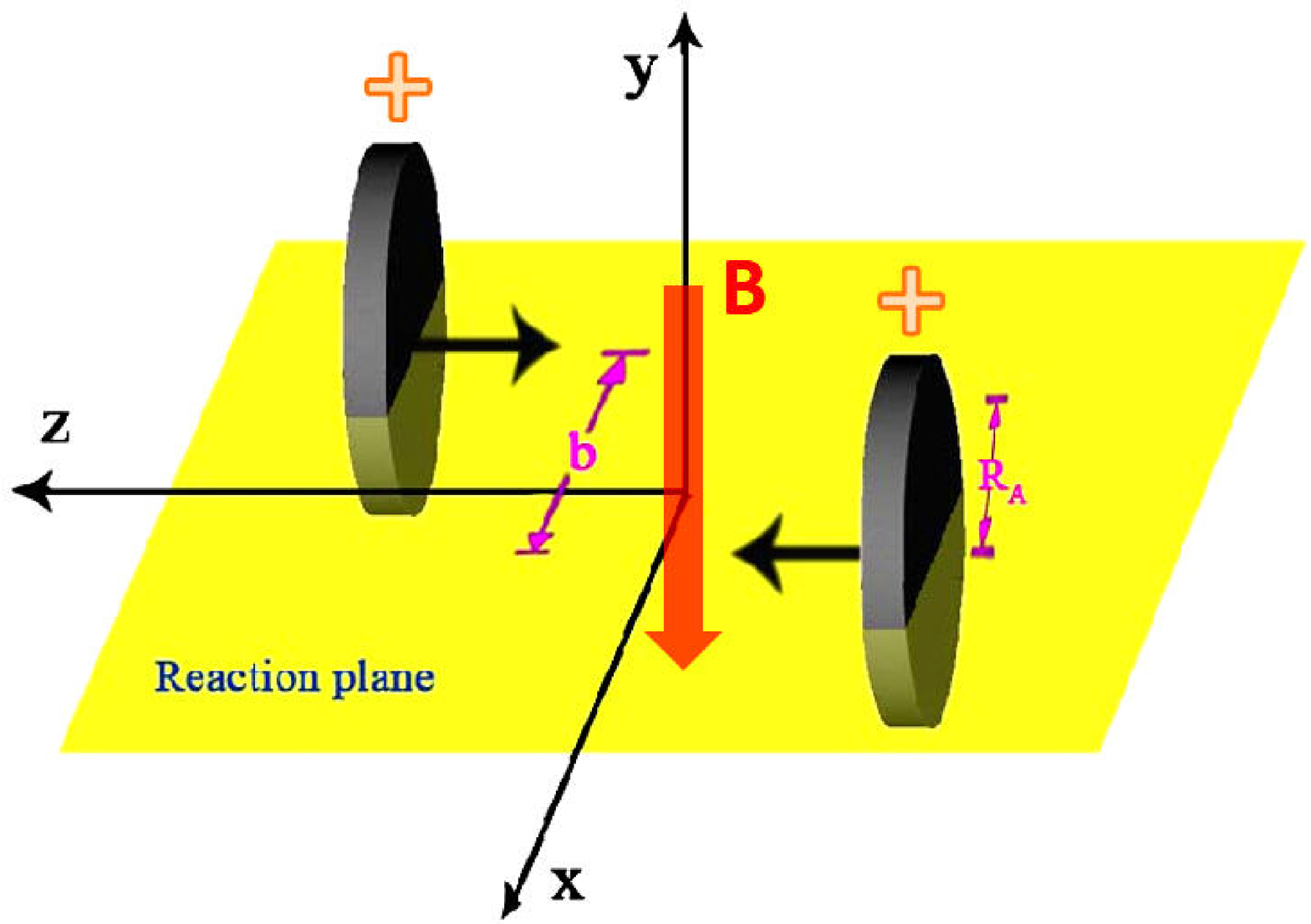}
\includegraphics[width=6.0cm]{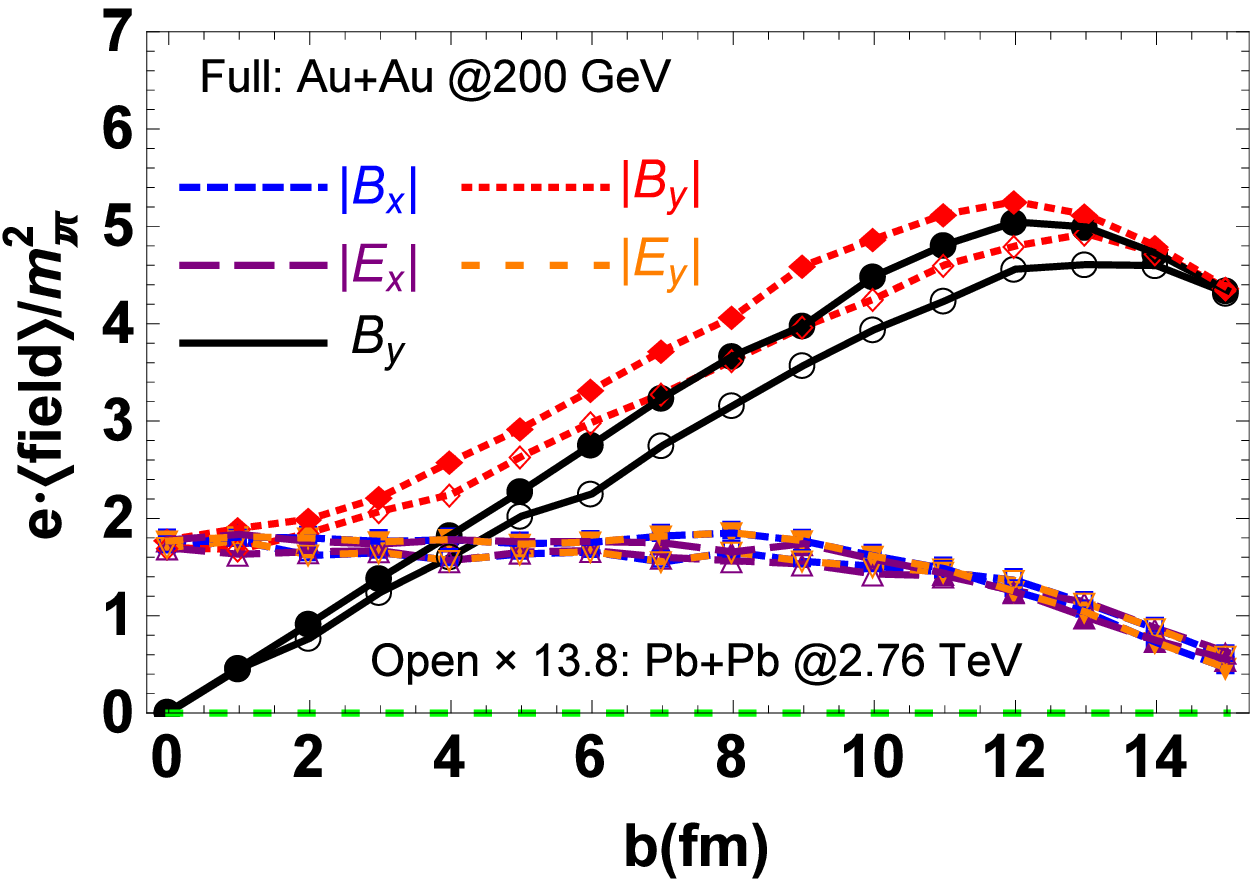}
\caption{(Left) The geometrical illustration of a typical non-central heavy-ion collision. The magnetic field $\bB$ is expected to be perpendicular to the reaction plane. (Right) The electromagnetic fields at $t=0$ and $\br={\vec 0}$ versus the impact parameter $b$. Figures are modified from Ref.~\cite{Deng:2012pc,Huang:2015oca}.}
\label{coll_geo}
\end{center}
\vspace{-0.7cm}
\end{figure}

In \fig{coll_geo} (left), we illustrate a typical non-central heavy-ion collision event. Two nuclei are accelerated to very high speed and then collide. During the collision, the two nuclei produce two counter-propagating currents which in turn induce a magnetic field whose direction is perpendicular to the reaction plane. (Strictly speaking, only after the event average the direction of the magnetic field is perpendicular to the reaction plane. On the event-by-event basis, the direction of the magnetic field fluctuates as a consequence of the fluctuation of the proton position in the nucleus~\cite{Bloczynski:2012en,Bloczynski:2013mca}.) The question is: how strong the magnetic field can be? We can make a rough estimate by using the Biot-Savart formula, $eB\sim \g \a_{\rm EM} Z/ R_A^2$, where $\g$ is the Lorentz gamma factor associated with the moving nuclei, $\a_{\rm EM}$ is the fine structure constant, $Z$ is the charge number of the nucleus, and $R_A$ is the radius of the nucleus. For RHIC Au + Au collisions at $\sqrt{s}=200$ GeV, the magnetic field is very large, $\sim 10^{19}$ Gauss, and for LHC Pb + Pb collisions at $\sqrt{s}=2.76$ TeV the magnetic field can reach the order of $10^{20}$ Gauss.

More realistic numerical simulations indeed confirm that the high-energy heavy-ion collisions can generate extremely strong magnetic fields as well electric fields. In \fig{coll_geo} (right), we present the numerical result for the magnetic fields as well as the electric fields at $t=0$ (defined as the time when the two nuclei overlap maximally) and at $\br=\vec 0$ (the center of the overlapping zone)~\cite{Deng:2012pc,Huang:2015oca}. The simulations show some interesting features of the electromagnetic fields in heavy-ion collisions. (1) When $b<2R_A$, $\lan e\vec B\ran$ ($\lan\cdots\ran$ means event average) is perpendicular to the reaction plane and its strength is roughly linearly proportional to $b$. (2) The strength $e|\vec B|$ is roughly proportional to the beam energy $\sqrt{s}$. (3) The event-by-event fluctuation of the proton position induces a finite electric field at the center $\br=\vec 0$ which after the event average goes to zero. But in some asymmetric collisions, e.g., the Cu + Au collision, there is a persistent electric field whose strength is of the same order as the magnetic field~\cite{Hirono:2012rt,Deng:2014uja}. (4) On event-by-event basis, the direction of the magnetic field fluctuates which leads to a suppression in the correlation between the azimuthal angle $\Psi_B$ and the second harmonic plane angle like the participant plane angle $\Psi_2$.  (5) The initial magnetic field is dominated by the spectator protons, the participants and the hot quarks do not contribute much. This causes a very fast decay of the magnetic field as a consequence of the fast fly-away of the spectators from the collision zone. However, if the produced matter is a conductor, the Faraday effect may retain the magnetic field in the medium for a much longer while~\cite{Tuchin:2010vs}.

The main message here is that there exist strong transient electromagnetic fields in high-energy heavy-ion collisions which are actually the strongest fields that we have ever known in our current universe. The strength of such magnetic fields is much larger than the current masses squared of light quarks, $m_u^2, m_d^2$, and thus is capable of inducing significant quantum effects in the QGP.

\subsection {Vorticity}\label{sec:vor}
\begin{figure*}
\begin{center}
\includegraphics[width=6cm]{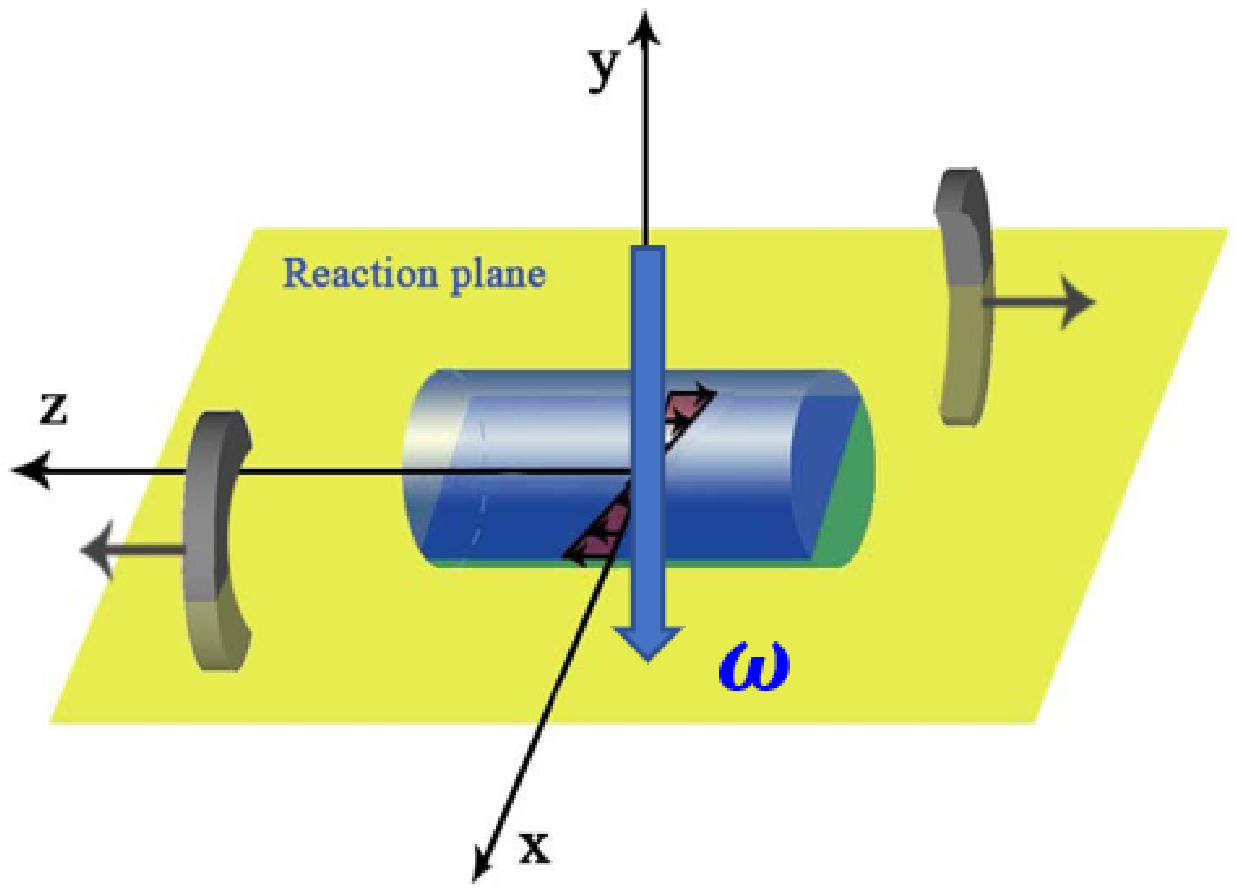}
\includegraphics[width=6cm]{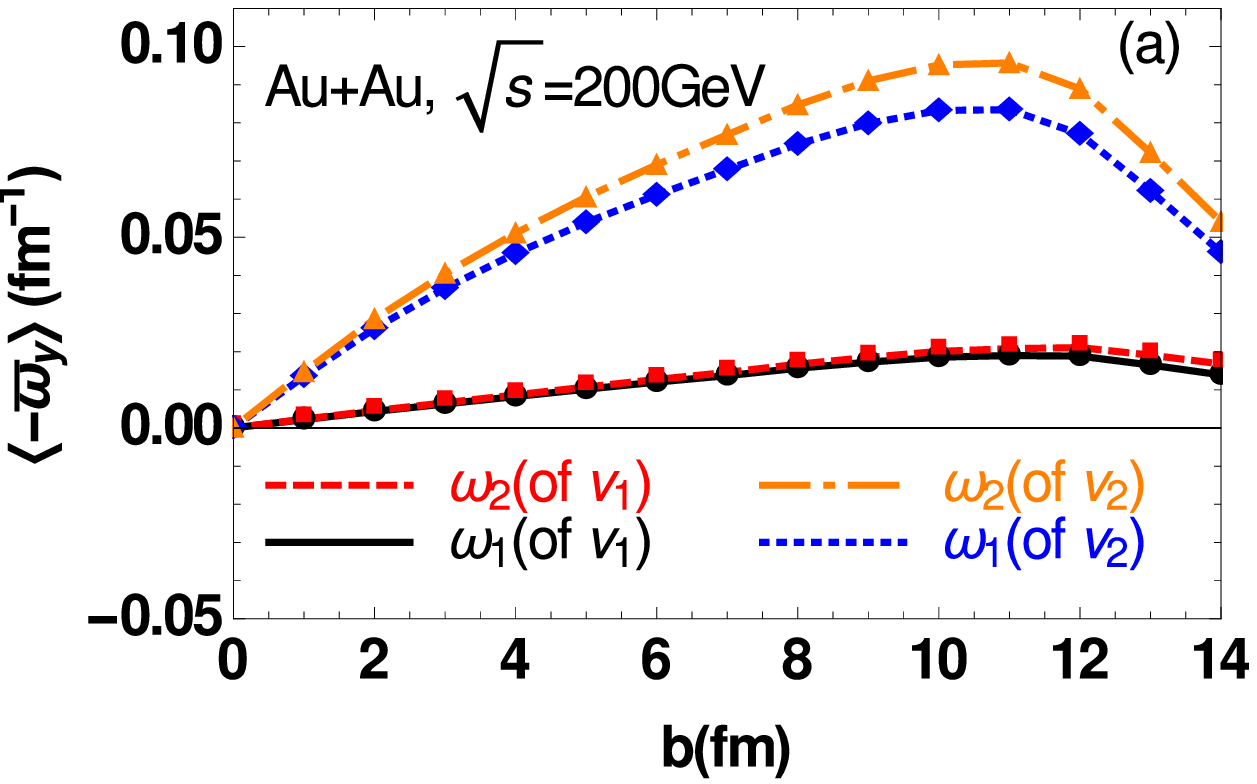}
\caption{(Left) The illustration of the appearance of the flow vorticity in a non-central heavy-ion collision. (Right) The vorticity at $\t=\t_0$ (the initial proper time) and $\w=0$ averaged over the overlapping region and then averaged over events for RHIC Au + Au collisions at $\sqrt{s}=200$ GeV. Only the $y$-component of the vorticity is sizable, other components are negligibly small. Different curves correspond to different definitions of vorticity and velocity, see Ref.~\cite{Deng:2016gyh} for details.}
\label{fig-vor-b-200}
\end{center}
\end{figure*}

The existence of strong magnetic fields in heavy-ion collisions suggests the existence of fast rotation and/or large flow vorticity in the QGP. In classical physics, the Larmor's theorem relates the motion of a charged particle in a magnetic field $\bB$ to the motion in a rotating frame with angular velocity $\vec \O=-q\bB/(2m)$ with $m$ and $q$ the mass and charge of the particle. In quantum physics, the Einstein-de-Haas effect is the generation of a rotation by magnetizing a ferromagnetic material which is a consequence of the spin polarization by the magnetic field. Besides, it is also natural to expect that heavy-ion collisions could create flow vorticity. Let us consider a non-central heavy-ion collision with impact parameter $b$ as shown in \fig{fig-vor-b-200} (left). Before the collision, the total angular momentum $J_0$ with respect to the collision center is roughly given by $Ab\sqrt{s}/2$ with $A$ the total nucleon number in one nucleus. After the collision, a fraction of $J_0$ will be kept in the produced quark-gluon matter and manifests itself as a shear of the longitudinal flow because the equation of state of the quark-gluon matter is too soft to support a global rigid rotation. Thus the nonzero local vorticity arises as a consequence of this shear flow and the direction of the vorticity would be merely along the direction of angular momentum which is perpendicular to the reaction plane.

We have performed more realistic numerical simulations recently~\cite{Deng:2016gyh,Deng:2016yru}; see also ~\cite{Jiang:2016woz}. In \fig{fig-vor-b-200} (right), we show the vorticity as a function of the impact parameter for RHIC Au + Au collisions at $\sqrt{s}=200$ GeV. The vorticity is defined as ${\vec\o}_{1}={\vec\nabla}\times\vec v$ and ${\vec\o}_{2}=\g^2{\vec\nabla}\times\vec v$ where $\vec v_1$ represents the velocity of particle flow while $\vec v_2$ represents the velocity of energy flow. According to our simulation, the vorticity generated in the heavy-ion collisions is indeed very large and it shows up some interesting features. (1) After suitably averaged over the collision region and then over many events, the vorticity is perpendicular to the reaction plane. But both the magnitude and the azimuthal direction of the vorticity suffer from the event-by-event fluctuation, and as a consequence, the vorticity direction is blurred from being perfectly perpendicular to the reaction plane. (2) The vorticity is larger for larger impact parameter $b$ when $b\lesssim2R_A$ and when $b>2R_A$ it drops. (3) Although the total angular momentum $J_0$ increases with $\sqrt{s}$, the event-averaged vorticity at mid rapidity decreases with increasing $\sqrt{s}$. This counterintuitive behavior is mainly because that the quark-gluon medium at mid rapidity behaves closer to the Bjorken boost invariant picture at higher $\sqrt{s}$ and thus supports lower vorticity.

The vorticity can be measured through measuring the spin polarization of hadrons, e.g., the $\L$ hyperon~\cite{Liang:2004ph,Becattini:2007sr,Huang:2011ru,Pang:2016igs}. The underlying mechanism is the spin-orbit coupling. The vorticity induces orbital motion of the quarks which then polarizes the spin of quarks through spin-orbit coupling. The spin polarization of quarks can be inherited by hadrons of finite spin, e.g., the $\L$ hyperon, after hadronization, which can in turn be measured. Recently, the STAR collaboration at RHIC measured the $\L$ and $\bar{\L}$ spin polarization with respect to the reaction plane from which they deduced an averaged vorticity over the beam energy region $7.7$ GeV$<\sqrt{s}<$200 GeV: $\o\sim (9\pm 1)\times 10^{21}/s$~\cite{STAR:2017ckg}. This is the highest vorticity comparing to the vorticity of all other known fluids.

\section{Anomalous chiral transports}
\label{sec-ano}
Perhaps the most intriguing effects that the strong electromagnetic fields and the strong vorticity can induce are the anomalous chiral transports (ACTs). The famous examples of ACTs are the chiral magnetic effect, chiral vortical effects, chiral separation effects, etc. We in this section will give a pedagogical discussion about the underlying mechanisms of the ACTs; more discussions can be found in Refs.~\cite{Kharzeev:2015znc,Huang:2015oca,Hattori:2016emy}

\begin{figure*}
\begin{center}
\includegraphics[width=5.3cm]{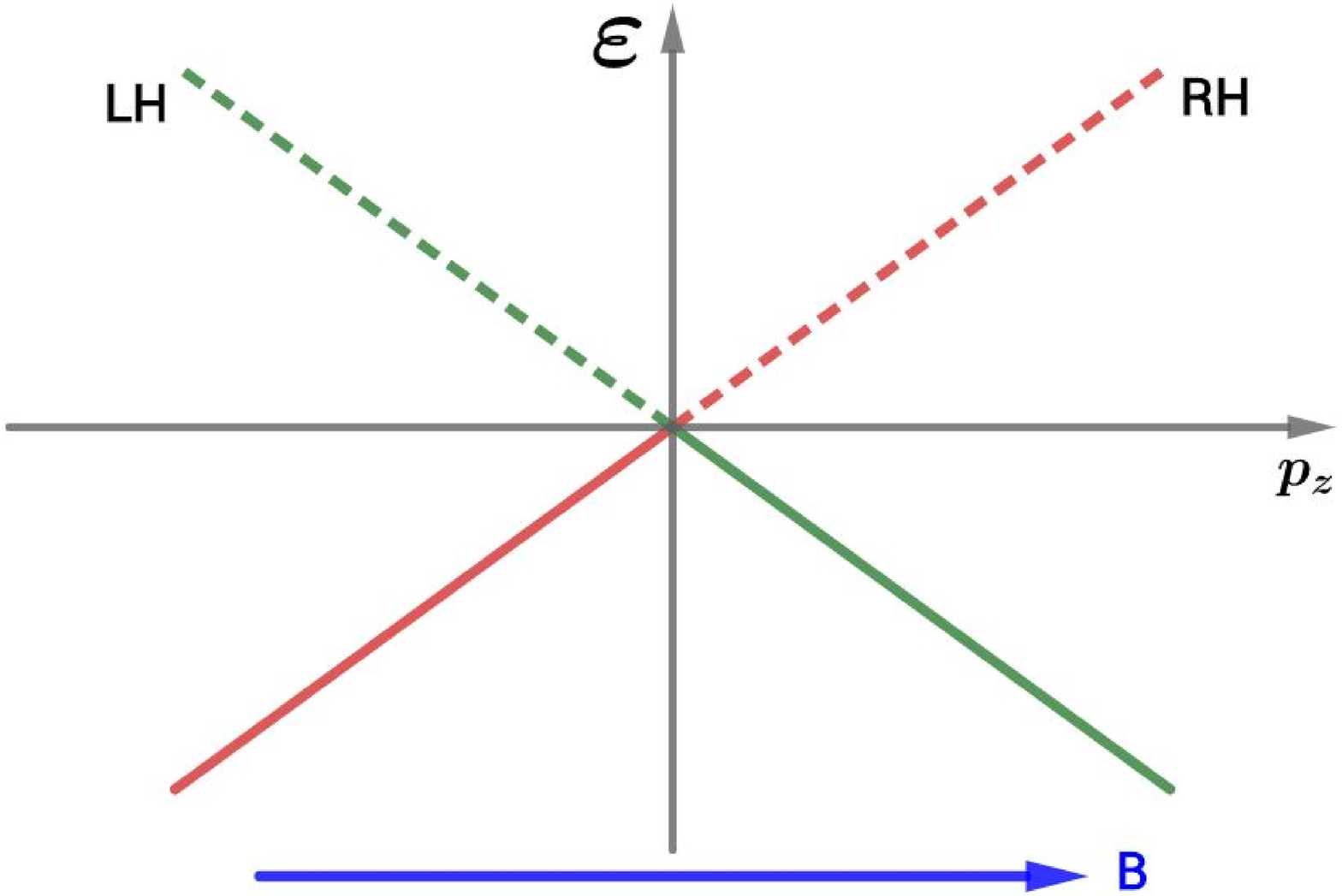}
\includegraphics[width=5.0cm]{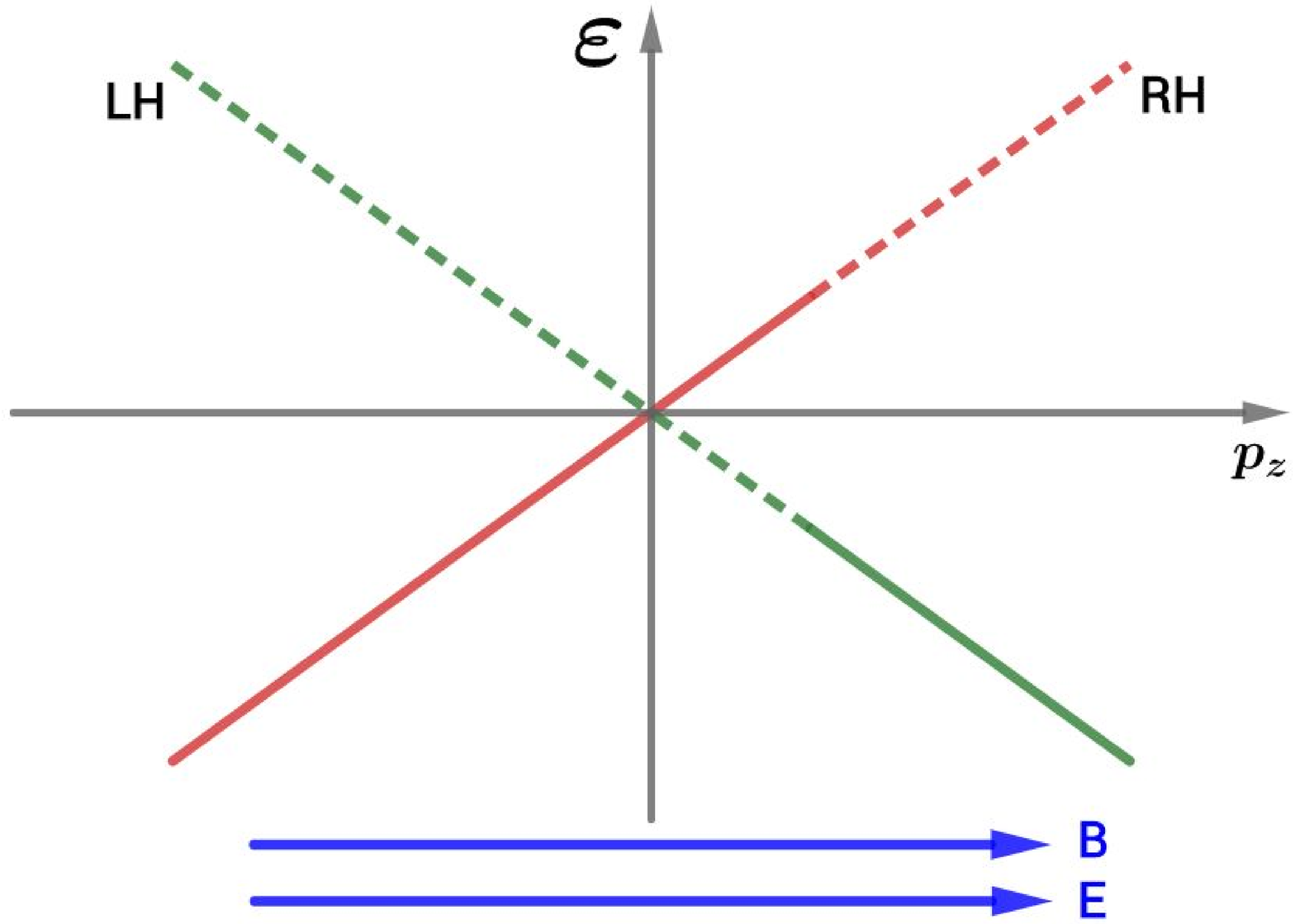}
\caption{(Left) Lowest Landau level in a strong magnetic field. (Right) The electric field pumps LH fermions into RH fermions near the zero node which leads to the chiral anomaly.}
\label{fig-ca}
\end{center}
\end{figure*}
Let us start with the $U_A(1)$ chiral anomaly which is the basis of the ACTs. Let us consider a massless Dirac fermion of charge $e>0$ in a strong constant magnetic field along the $z$ direction. This is the usual Landau problem in quantum mechanics. The energy spectrum consists of different Landau levels,
\begin{eqnarray}
\mathcal{E}_n^2=p_z^2+2neB, \;\; n=0,1,2,\cdots,
\end{eqnarray}
where $n$ labels the Landau levels. The lowest Landau level (LLL), $n=0$, is special, see \fig{fig-ca} (left). First, the LLL is gapless while all the higher Landau levels are gapped with a magnitude of $\sqrt{2neB}$. Thus when $eB$ is large, we need to consider only the LLL. Second, the spin of LLL is fully polarized, namely, the LLL is non-degenerate in spin. All the states of the LLL are of spin up. In a many-body picture, this means that the LLL fermions are all have spin up. Third, the dynamics of LLL is 1+1 dimensional because the transverse motion of the LLL fermion is frozen. We define the chirality for each LLL fermion according to its momentum direction with respect to its spin direction. If $p_z$ is parallel to the spin, we call it a right-handed (RH) fermion; if $p_z$ is opposite to its spin, we call it a left-handed (LH) fermion.

In this situation, the numbers of RH and LH fermions are separately conserved, namely, $\pt_\mu J^\mu_{R/L}=0$ with $J^\m_{R/L}=(1/2)\jb\g^\m(1\pm\g_5)\j$, or equivalently, $\pt_\mu J^\mu_V=0$ and $\pt_\mu J^\mu_A=0$, where the vector and axial currents are defined as $J_V^\m=J_R^\mu+ J_L^\mu$ and $J^\mu_A=J^\m_R-J^\m_L$.

Now we apply an electric field in the same direction as the magnetic field, see \fig{fig-ca} (right). Near the node $p_z=0$, the electric field can drive the downward moving particles to upward moving and thus tune some LH fermions to RH fermions. In such case, although the sum of the numbers of RH and LH fermions, $N_V=N_R+N_L$, is still conserved, the difference $N_A=N_R-N_L$ is not conserved. We can calculate the changing rate of $N_A$ in the following way. Suppose the RH and LH fermions have Fermi surfaces $p_F^{R/L}$. Then
\begin{eqnarray}
N_{R/L}=V\frac{p_F^{R/L}}{2\p}\frac{eB}{2\p},
\end{eqnarray}
where $eB/(2\p)$ is the transverse density of state and $V$ is the spatial volume. Thus,
\begin{eqnarray}
\frac{dN_{R/L}}{dt}=V\frac{\dot{p}_F^{R/L}}{2\p}\frac{eB}{2\p}=\pm V\frac{eE}{2\p}\frac{eB}{2\p},
\end{eqnarray}
which implies $d N_V/dt=0$ and $d N_A/dt=Ve^2EB/(2\p^2)$ which, written in a differential form, is essentially
\begin{eqnarray}
\label{chiralanom}
\pt_\mu J_A^\m=\frac{e^2}{2\p^2}\bE\cdot \bB.
\end{eqnarray}
This result is know as chiral or axial anomaly. Although we obtain \eq{chiralanom} by considering strong magnetic field so that only the LLL is occupied, the result is actually true for weak magnetic field too, because the higher Landau levels are degenerate in chirality so that they do not contribute to \eq{chiralanom}.
\begin{figure*}
\begin{center}
\includegraphics[width=6cm]{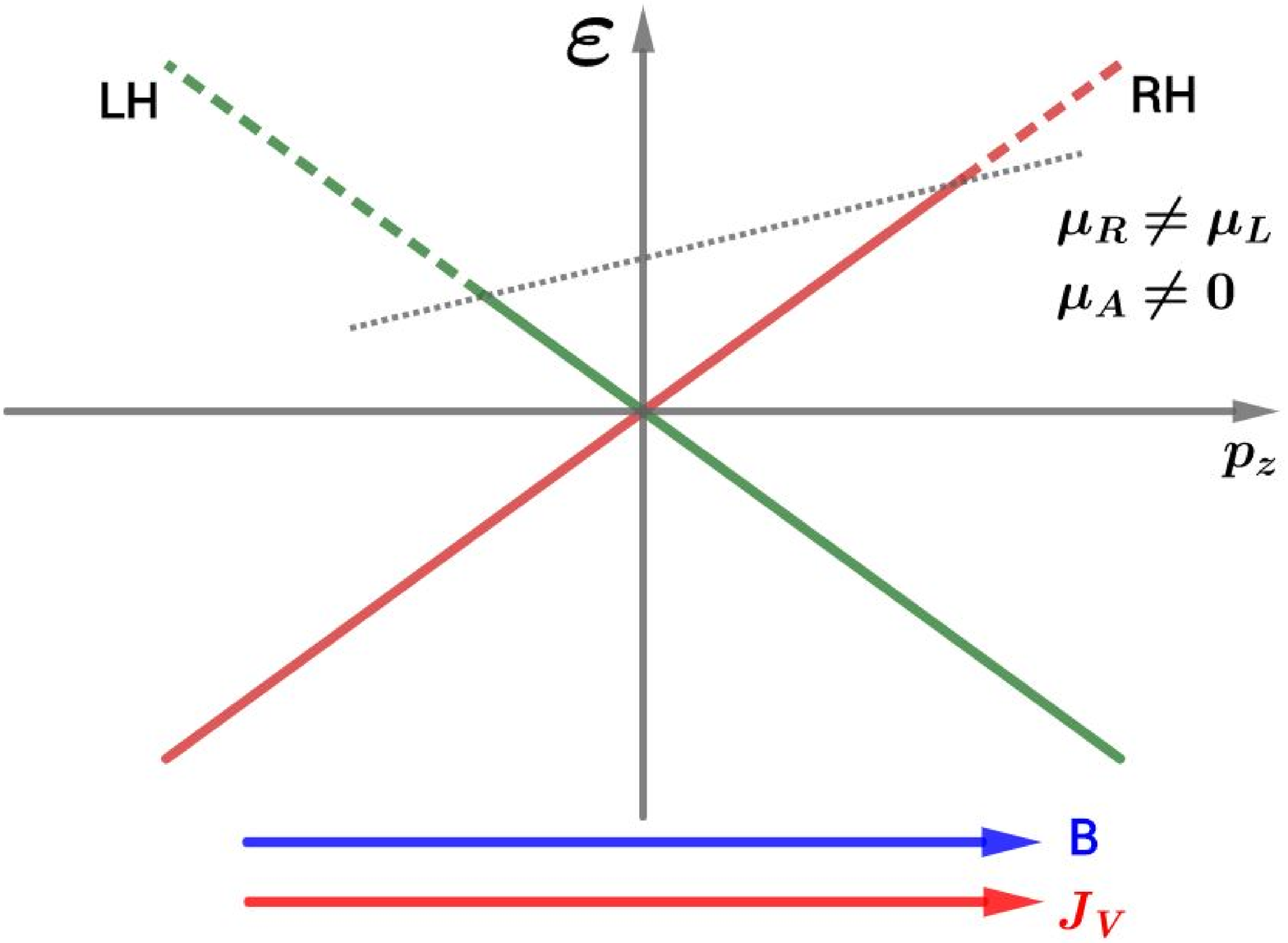}
\includegraphics[width=6cm]{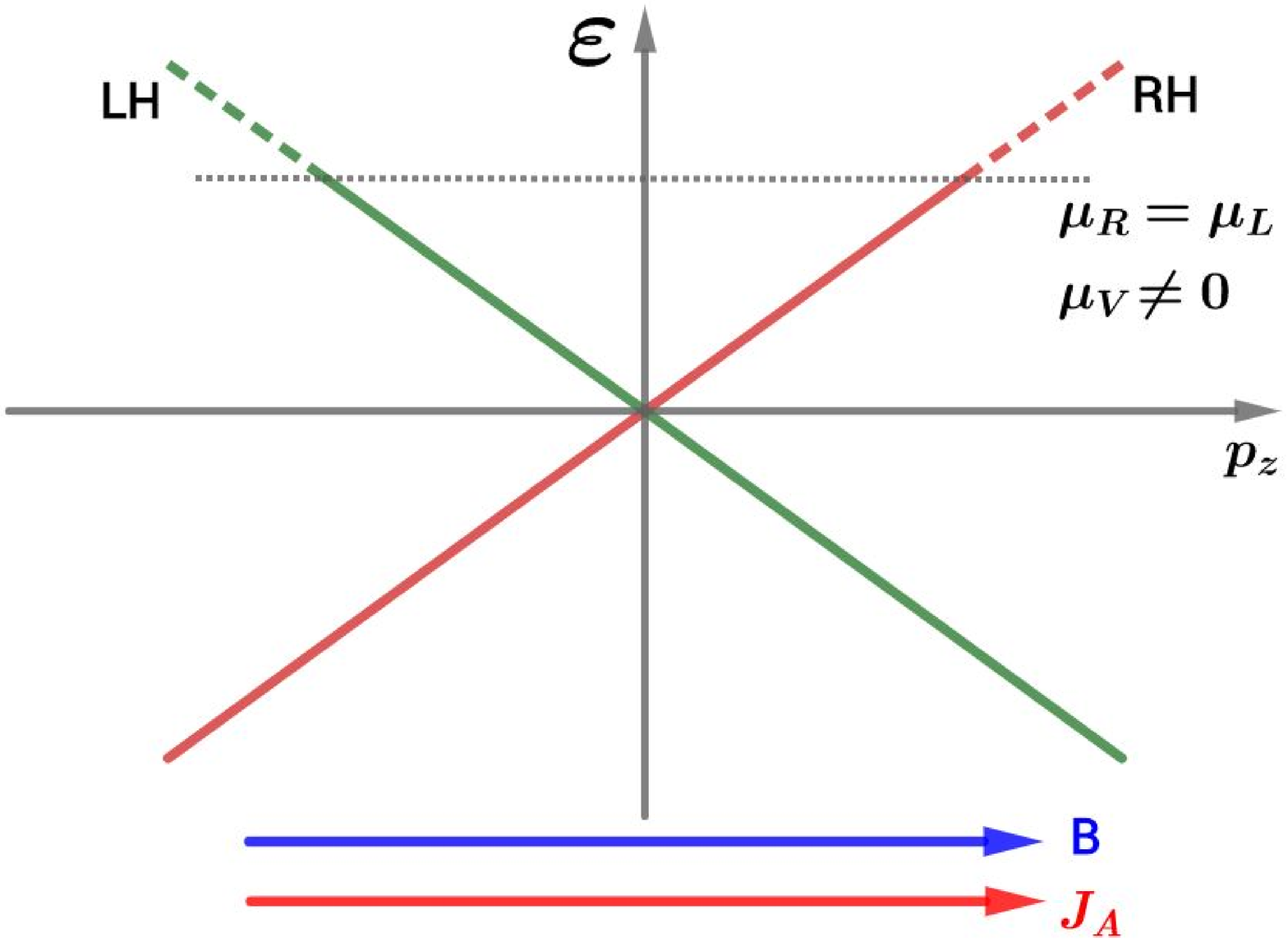}
\caption{(Left) The arising of the chiral magnetic effect. (Right) The arising of the chiral separation effect.}
\label{fig-ckt}
\end{center}
\end{figure*}

Now we remove the electric field and calculate the RH and LH currents along the magnetic field, see \fig{fig-ckt} (left). A current is equal to the density that the current carries times the velocity of the constitute particles. For massless particle, the velocity is one. Thus we have
\begin{eqnarray}
\label{cmecur1}
J_{R/L}=\pm n_{R/L}=\pm\frac{p_F^{R/L}}{2\p}\frac{eB}{2\p},
\end{eqnarray}
where the minus sign for LH current is because LH fermions move opposite to the direction of the magnetic field. This gives
\begin{eqnarray}
\label{cmecur2}
J_{V}=\frac{p_F^{R}-p_F^L}{2\p}\frac{eB}{2\p}=\frac{\m_A}{2\p^2}eB,
\end{eqnarray}
where we have defined the axial chemical potential $\m_A=(\m_R-\m_L)/2$. This current is the chiral magnetic current~\cite{Kharzeev:2007jp,Fukushima:2008xe}. Similarly,
\begin{eqnarray}
\label{csecu}
J_{A}=\frac{p_F^{R}+p_F^L}{2\p}\frac{eB}{2\p}=\frac{\m_V}{2\p^2}eB,
\end{eqnarray}
where $\m_V=(\m_R+\m_L)/2$. This is the chiral separation effect (CSE)~\cite{Son:2004tq}; it appear even when $p_F^R=P_F^L$, see \fig{fig-ckt} (right). We note that the CME has very intriguing properties. First, it is a macroscopic quantum effect. Second, its occurrence requires P and CP violation in the medium. Third, the generation of the CME current is time reversal even, namely, there is no associated entropy production. Thus the CME current is a kind of superconducting current. We also emphasize that the CME conductivity is fixed by the chiral anomaly and thus is free of renormalization.

In classical physics, a charged particle of mass $m$ in a magnetic field is equivalent to being in a rotating frame with frequency $eB/(2m)$. This suggests that there may exist effects analogous to CME and CSE but induced by rotation or vorticity. Let us consider a massless particle in a rotating frame. The particle feels a Coriolis force, $\vec F=2p \dot{\vec x}\times\vec\o + O(\o^2)$, where $\vec\o$ is the rotating frequency. We have assumed that $\o$ is so small that we neglect the centrifugal force which is $O(\o^2)$. As the Coriolis force is very similar with the Lorentz force (replacing $eB$ by $2p\o$), we can consider the ``Landau level problem'' in rotating frame. Let us again consider only the LLL. Comparing to the magnetic case, the only difference is that the expression for the density is modified: $n_{R/L}=(2\p)^{-2}\int_0^\infty dp_z 2p_z\o \h(p_F^{R/L}-p_z)=(p_F^{R/L})^2\o/(2\p)^2$. Thus the currents:
\begin{eqnarray}
\label{cvecu1}
J_{V}&=&n_R-n_L=\frac{\m_V\m_A}{\p^2}\o,\\
J_{A}&=&n_R+n_L=\frac{\m_V^2+\m_A^2}{2\p^2}\o.
\end{eqnarray}
This is the chiral vortical effects~\cite{Erdmenger:2008rm,Banerjee:2008th,Son:2009tf}. More rigorous derivation shows that there is an additional term, $T^2\o/6$, in $J_A$ which may be related to the gravitational anomaly~\cite{Landsteiner:2011cp}.

The electric field can also lead to anomalous transport, that is the chiral electric separation effect (CESE)~\cite{Huang:2013iia}. It is not directly related to the chiral anomaly and its appearance requires both P and C violation. The CESE represents an axial current along the direction of the electric field. Its expression for two flavor QCD up to leading-log accuracy is given by~\cite{Jiang:2014ura}
\begin{eqnarray}
{\vec J}_A\approx 14.5163\Tr (Q_e Q_A)\frac{\m_V\m_A}{T^2} \frac{eT}{g^4\ln (1/g)}\vec E,
\end{eqnarray}
where $Q_e$ and $Q_A$ are charge matrix and axial matrix in flavor space, $g$ is the strong coupling constant.

There can emerge collective modes from the coupled evolution of the axial and vector charges via CME and CSE, or vector CVE and axial CVE, or CESE and the usual Ohm's law. For example, the continuation equations for vector and axial charges can be written in terms of RH and LH charges
\begin{eqnarray}
\label{continuation}
\pt_t J^0_{R/L}+{\vec\nabla}\cdot\bJ_{R/L} &=&0.
\end{eqnarray}
Now we substitute the CME and CSE expressions, and consider small departures from equilibrium for $J^0_{R,L}$ and $\m_{R,L}$ and keep linear terms in the small departures:
\begin{eqnarray}
\label{waveequ}
\pt_t\d J^0_R+\frac{e^2}{4\p^2\c_R}\bB\cdot\vec\nabla\d J^0_R &=&0,\\
\label{waveequ2}
\pt_t\d J^0_L-\frac{e^2}{4\p^2\c_L}\bB\cdot\vec\nabla\d J^0_L &=&0,
\end{eqnarray}
where $\c_R=\pt J^0_R/\pt\m_R$ and $\c_L=\pt J^0_L/\pt\m_L$ are susceptibilities for RH and LH chiralities. These two equations express two collective, gapless, wave modes which are called the chiral magnetic waves (CMWs)~\cite{Kharzeev:2010gd}. Similarly, if we consider the CESE and the Ohm's law, we can find new collective modes, the chiral electric waves and axial or vector density waves~\cite{Huang:2013iia}; if we consider the two CVEs, we can find chiral vortical waves~\cite{Jiang:2015cva}. Further more, if one consider, e.g. both magnetic field and vorticity, there emerge more collective modes. Finally, we summarize the anomalous chiral transports in Table (\ref{tab-1}).

\begin{table}
\centering
\caption{Table of anomalous chiral transports}
\label{tab-1}       
\begin{tabular}{|c|c|c|c|}
\hline
  & $\vec E$ & $\vec B$ & $\vec \o$  \\\hline
$\displaystyle\vec J_V$ & $\displaystyle\s$ & $\displaystyle\frac{e\m_A}{2\p^2}$ & $\displaystyle\frac{\m_V\m_A}{\p^2}$ \\\hline
$\displaystyle\vec J_A$ & $\displaystyle\propto\frac{\m_V\m_A}{T^2}\s$ & $\displaystyle\frac{e\m_V}{2\p^2}$ & $\displaystyle\frac{T^2}{6}+\frac{\m_V^2+\m_A^2}{2\p^2}$ \\\hline
Collective modes: & chiral electric wave & chiral magnetic wave & chiral vortical wave \\\hline
\end{tabular}
\end{table}

\section{Anomalous chiral transports in heavy-ion collisions}
\label{sec-hic}
If the ACTs occur in the heavy-ion collisions, what will be their experimental signals? As the magnetic fields are, on average over events, perpendicular to the reaction plane, the CME will drive a charge dipole in the fireball which will be converted to a charge dipole in momentum space and can be tested via detecting the final charged hadron distribution. Similarly, the CVE will induce a baryonic dipole with respect to the reaction plane and can be tested by detecting the final baryon number distribution. The CESE, together with the CME, may drive a charge quadrupole in, e.g. Cu + Au collisions and thus may be tested. The CMW may induce a splitting between the elliptic flows of positively and negatively charged particles and the CVW may induce a splitting between the elliptic flows of baryons and antibaryons. We will focus on the CME.

Because $\m_A$ is fluctuating over events, the CME-induced charged dipole is also fluctuating. To measure this dipolar fluctuation, one can use the following charge-dependent two-particle azimuthal correlation~\cite{Voloshin:2004vk}:
\begin{eqnarray}
\label{gamma}
\g_{\a\b}&\equiv&\lan\cos(\f_\a+\f_\b-2\J_\rp)\ran,
\end{eqnarray}
where $\a$ and $\b$ denote the charges, $\f_\a$ and $\f_\b$ are the corresponding azimuthal angles, $\J_\rp$ is the reaction plane angle, and $\lan\cdots\ran$ is event average. It is easy to see that a charge separation with respect to the reaction plane will give a positive $\g_{+-}$ and $\g_{-+}$ (will be referred to as $\g_{\rm OS}$) while a negative $\g_{++}$ or $\g_{--}$ (will be referred to as $\g_{\rm SS}$). In real experiments, the measurements are done with three-particle correlations where the third hadron (of arbitrary charge) is used to reconstruct the reaction plane.

The correlation $\g_{\a\b}$ was measured by STAR Collaboration at RHIC for Au + Au and Cu + Cu collisions at $\sqrt{s}=200$ GeV~\cite{Abelev:2009ac,Abelev:2009ad}, by PHENIX Collaboration at RHIC for Au + Au collisions at $\sqrt{s}=200$ GeV~\cite{Ajitanand:2010rc}, by ALICE Collaboration at LHC for for Pb + Pb collisions at $\sqrt{s}=2.76$ TeV~\cite{Abelev:2012pa}, and by CMS Collaboration at LHC for Pb+Pb collisions at $\sqrt{s}=5.02$ TeV~\cite{Khachatryan:2016got}. At mid-central collisions, these measurements show clear positive $\g_{\rm OS}$ and negative $\g_{\rm SS}$ with features in consistence with the expectation of the CME. The STAR Collaboration also measured $\g_{\a\b}$ at different beam energies~\cite{Adamczyk:2014mzf} and found that $\g_{\a\b}$ persists as long as the beam energy is larger than $19.6$ GeV.

However, the interpretation of the data is ambiguous owing to possible background effects that are not correlated to CME. For example, the transverse momentum conservation (TMC) can induce a back-to-back correlation which contributes to $\g_{\a\b}$~\cite{Pratt:2010gy,Pratt:2010zn,Bzdak:2010fd}, which can be subtracted by making a difference $\D\g\equiv\g_{\rm OS}-\g_{\rm SS}$ as the TMC is charge blind. There are also charge dependent background effects, like the local charge conservation (LCC) or neutral cluster decay~\cite{Schlichting:2010qia,Wang:2009kd} which gives a finite contribution to $\D\g$: $\D\g^{\rm LCC}\approx M v_2/N$
where $M$ is the number of hadrons in a local neutral cell, $N$ is the multiplicity, and $v_2$ is the elliptic flow. To disentangle the possible CME signal which is proportional to the magnetic field squared and the flow-related backgrounds which is proportional to $v_2$, one can either vary the backgrounds with the signal fixed or vary the signal with the backgrounds fixed. The former approach was carried out by using central U + U collisions~\cite{Voloshin:2010ut} where sizable $v_2$ can appear due to the prolate shape of uranium nucleus but the magnetic field should be small~\cite{Bloczynski:2013mca}. The experimental result from the STAR Collaboration shows that in the $0-1\%$ most central events of U + U collisions at $\sqrt{s}=193$ GeV there is indeed sizable $v_2$ and very small $\D\g$~\cite{Wang:2012qs}. However, it was found later that the total multiplicity is far less correlated to the number of binary collisions than expected~\cite{Adamczyk:2015obl}, thus the shape selection is hard to control. The latter approach can be achieved by using the isobaric collisions, such as $^{96}_{44}$Ru and $^{96}_{40}$Zr~\cite{Voloshin:2010ut,Deng:2016knn,Skokov:2016yrj,Huang:2017azw}. It is expected that Ru + Ru and Zr + Zr collisions at the same beam energy and same centrality will produce similar elliptic flow but a $10\%$ difference in the magnetic fields. Thus if the $\g$ correlation contains contribution from CME, the isobaric collisions could have a chance to see it by comparing the same $\g$ correlation in Ru + Ru and Zr + Zr collisions. The detailed simulation in Ref~\cite{Deng:2016knn} indeed found that the two collisions at $\sqrt{s}=200$ GeV have more than $10\%$ difference in the CME signal and less than $2\%$ difference in the elliptic-flow-driven backgrounds for the centrality range of $20-60\%$ by assuming that the CME contribution to the $\g$ correlation is $1/3$. Such a difference is feasible in current of heavy-ion collisions and RHIC have planed the isobaric run in 2018, so let us look forward to their result which will be a very important step towards the detection of the CME in heavy-ion collisions.

Recently, the CMS Collaboration reported the result of $\g_{\a\b}$ in high multiplicity events in p + Pb and Pb + Pb collisions at $\sqrt{s}=5.02$ TeV ~\cite{Khachatryan:2016got}, in which they found similar $\g_{\a\b}$'s in p + Pb and in Pb + Pb as functions of multiplicity. These results suggest that the gamma correlation at $5.02$ TeV or above in both p + Pb and Pb + Pb collisions are dominated by background effects. This is somehow compatible with the measurement of STAR collaboration in which the so-called $H$ correlation in measured at various energies through the Beam Energy Scan (BES) program~\cite{Adamczyk:2014mzf}. If one extrapolates the $H$ correlation from lower energies to $\sqrt{s}=5.02$ TeV, one can find that $H$ correlation is actually small there. See relevant articles in this volume. More studies are definitely needed.

To end this section, we discuss the possible observables for other ACTs. The CESE may lead to observable effect in, e.g., Cu + Au collisions where persist electric fields from Au to Cu nuclei exist~\cite{Ma:2015isa}. The CVE can induce baryon number separation with respect to the reaction plane~\cite{Kharzeev:2010gr} and CVW may cause a $v_2$ splitting between $\L$ and $\bar{\L}$~\cite{Jiang:2015cva}. Recently, the STAR Collaboration has reported the first measurement of CVE in Au + Au collisions at $\sqrt{s}=200$ GeV~\cite{Zhao:2014aja} and the data is consistent with the expectation of CVE. But, similar with the case of CME and CMW detections, the data contains contributions from background effects like the transverse momentum conservation and the local baryon number conservation, and we need more efforts to understand the data.

{\bf Acknowledgments:} This work is supported by NSFC with Grants No. 11535012 and No. 11675041 and the One Thousand Young Talents Program of China.

\end{document}